\documentclass[%
notitlepage,
twocolumn,
 amsmath,amssymb,
]{revtex4-1}

\usepackage{graphicx}
\usepackage{dcolumn}
\usepackage{bm}
\usepackage{xcolor}
\usepackage{mathtools}
\usepackage{bbold}

\newcommand{\ket}[1]{| #1 \rangle}
\newcommand{\bra}[1]{\langle #1 |}

\newcommand{\neu}[1]{ {\color{black} #1 }}
\begin{document}

\preprint{APS/123-QED}

\title{Long-range interaction in an open boundary-driven Heisenberg spin lattice:\\ A far-from-equilibrium transition to ballistic transport}

\author{Manuel Katzer}
\email{manuel.katzer@physik.tu-berlin.de}
\author{Willy Knorr}
\author{Regina Finsterh\"olzl}
\author{Alexander Carmele}
\affiliation{Technische Universit\"at Berlin, Institut f\"ur Theoretische Physik, Nichtlineare Optik und Quantenelektronik, Hardenbergstra{\ss}e 36, 10623 Berlin, Germany}

\date{\today}

\begin{abstract}
We study an open Heisenberg XXZ spin chain with long-range Ising-type interaction, which is incoherently driven at its boundaries and therefore a far-from-equilibrium steady state current is induced. Quantum Monte Carlo techniques make a regime accessible, where the system size is large enough to examine the transition from a diffusive to a ballistic transport regime. We find that the\neu{chain lengths}for this transition are increasing with decreasing range of the Ising-type interactions between distant spins, while the strength of the incoherent driving does not have a relevant effect on the transport transition. The transition can be explained by the suppression of ferromagnetic domains at the edges of the chain.
\end{abstract}
\maketitle
\section{INTRODUCTION}\label{sec:introduction}
In the field of strongly correlated quantum many body physics, the Heisenberg XXZ chain \cite{Heisenberg1928} is an important model since it allows an analytical solution in the case of unitary, reversible and nearest-neighbour dynamics \cite{Bethe1931,Dupont2020}.  
Recently, there has been growing interest in spin chains with long-range interaction,\neu{examined in classical approaches~\cite{LuijtenBloete1997classicallongrange,Valenzuela2003classicallongrange,GONCALVES2005classicallongrange} as well as in quantum spin chains~\cite{duerrprl05,sandvikprl10,koffelprl2012,bachelardkastner2013closedanalyticlongrange,cevolanipra15,gongprb2016,gongprb2016_2,maghrebiprl2017,bermudezprb17,frerotprb2017,zunkovic2018,vanderstraetenprl,renprb20,Halimeh2017closedlongrangekastner,Cevolani2018closedchainlongrange,halimehprb2018,zauner-stauberpre2018,klossandbar_lev2019closedchainlongrangemeanfield,liu2019closedlongrange,Hermes2020closedlongrange2d}. In these systems the interaction}exceed the widely examined nearest neighbour case, since many physical interactions show a power-law range over distance, e.g. Coulomb-\cite{molmerrmp10}, dipole-dipole-\cite{lahaye2009,Yan2013} or even van der Waals interaction \cite{molmerrmp10}.\neu{In the field of closed quantum systems, it was possible to identify phase transitions dependent on the long-range parameter $\alpha$~\cite{klossandbar_lev2019closedchainlongrangemeanfield}, and for sufficiently small $\alpha$, analytical solutions with mean field techniques were found \cite{bachelardkastner2013closedanalyticlongrange}.}

\neu{There has been a multitude of proposals towards experimental realisations for the Heisenberg model, many of which where successfully implemented. In the context of transport properties, for instance, a chain of trapped atomic ions was recently proposed to examine the spin conductivity by inducing a source and a drain on two different sites~\cite{trautmannhauke2018}, which was later realized experimentally~\cite{maieretalhaukeprl}. In addition to this, there are numerous possible technological applications succesfully simulated with Heisenberg chains, ranging from CNOT-gates made of SQUIDS~\cite{Heule2011CNOT_SQUID}, to quantum batteries~\cite{Le2018quantumbattery} and novel transistors~\cite{Marchukov2016transistor}.}%

Following this and related recent experimental progress \cite{KinoshitaNature2006,hildprl2014,LangenScience2015,TangPRX2018}, it has become an emerging topic to describe the chain in an open quantum system approach, to account for dissipative-induced effects  \cite{jesenkoprb2011,LeviPRL2016,MevedyevaPRB2016,Monthus2017,savonapra2018,LjubotinaPRL2019,carmele2019non,su2013collective,gegg2018superradiant,carmele2015stretched,polettipra2019,finsterhoelzl2020nonequilibrium,kaestle2020memorycritical}.
A paradigm in this field of open quantum spin chain dynamics is the investigation of non-equilibrium, non-trivial steady-states and the resulting quantum transport properties. A common example is given with the boundary-driven spin chain \cite{SaitoEPL2003,WichterichPRE2007,ZnidaricJPA2010,znidaricprl11,prosenprl11,karevskiprl2013,PopkovPRE2013,LandiPRB2015,znidaricprl2016,prbleon17}, i.e. the spin chain is driven incoherently with a bias at its edges in a Lindblad-based approach \cite{lindbladoriginal,carmele2019non,breuer} and then examined in the long-time limit.
A variety of non-equilibrium phenomena has already been found in such systems, e.g. negative differential conductivity \cite{benentiprb09,BenentiEPL2009}, anomalous transport \cite{ZnidaricJSM2011,prosenprl2011_2,SchulzJSM2020}, and current-enhancement by dephasing \cite{ZnidaricNJP2010,MendozaArenasPRB2013,MendozaArenasJSM2013}. An extensive overview of the whole research field was recently provided \cite{bertiniARXIV2020}.%

So far, most studies on quantum transport phenomena in open spin chains have been focused to the interaction between directly neighbouring sites. 
Here, we are interested in the dynamics and the generated non-equilibrium steady-state in case of long-range interaction in a boundary-driven open system.%
We will show that the transport properties change qualitatively with increasing long-range interaction from diffusive to ballistic transport. Our study aims therefore to unravel the underlying physics in state-of-the-art experiments in which it is possible to control the range $1/r^\alpha$ of those long-range interactions, effectively adjusting the exponent $0 < \alpha < 3$ \cite{Britton2012,Islam2013,Richerme2014,Jurcevic2014,SmithNatPhys2016,Neyenhuise2017,jurcevicprl2017}. 
This work shows that one can obtain very special insights about the effect of spin-flips and Ising interaction on spin transport, when treating a long-range Heisenberg XXZ chain in an open quantum system approach by driving it at the boundaries.
Here, we extend a previous study \cite{prbleon17} which focuses on the robustness of the steady-state current against disorder in two extreme limits: the nearest-neighbour ($\alpha\to\infty$) and the strong long-range case ($\alpha\to0$). It was demonstrated that long-range Ising interaction stabilizes the transport in the chain against negative differential conductivity (NDC) effects. Those NDC effects had been found before for the nearest neighbour case \cite{benentiprb09}. It was furthermore shown that this robustness even holds for a disordered chain. 
In the following, we discuss the transition between ballistic and diffusive transport behaviour by investigating the intermediate $\alpha$ regime.
Dependent on the range of the $J_z$-interaction, mediated by the parameter $\alpha$, \neu{we find a chain-length dependent smooth transition,} where the transport changes from diffusive to ballistic behaviour, i.e. the NESS current saturates and no longer decreases with growing chain lengths. The effect is qualitatively independent of the driving strength, at least beyond the linear case of very weak driving. This is also interesting since the isotropic Heisenberg chain (with only nearest neighbour interaction) was initially expected to show ballistic transport due to its integrability, but later was shown to be superdiffusive \cite{znidaricprl11}. 

In this paper, we numerically demonstrate that if the Ising interaction has a longer range than nearest neighbour interaction, the ballistic transport regime can also be reached at the isotropic point where $J_x = J_z$, which in the nearest neighbour case is only possible for an anisotropic spin chain \cite{znidaricprl2016,LjubotinaNatCom2017}. Note that we refer to $J_z$ here as the sum of all Ising interactions of one site with all other sites.
The numerical method employed is based on the averaging of quantum trajectories \cite{Molmer93,plenioknightrmp98,daleyreview2014trajectories} and  enables a description of chain lengths around N=20, a regime which is not computationally accessible in the master equation framework.%

The paper is organized as follows: after the introduction, cf.~Sec.~\ref{sec:introduction}, we introduce in Sec.~\ref{modelsection} the model of a Heisenberg chain with long-range interaction and incoherent driving. We also choose the steady state spin current as the central figure of merit in the strong-driving limit beyond the linear regime. 
In Sec.~\ref{centralresultssection}, the main result of a transition to ballistic transport is presented, as well as the qualitative independence of this effect towards the driving (in-scattering rate $\Gamma$). In Sec.~\ref{discussionsection}, we give a physical explanation of the occurring transition and its strong dependence on the long-range parameter $\alpha$ by discussing the suppression of ferromagnetic domains in the chain and by showing how the Ising interaction alters the eigenvectors of the Hamiltonian already for a chain with $N=3$, while long-range leads to an out-cancelling of this current inhibiting effect of the Ising interaction. Further details on the eigenvectors, related polarisation profiles and on the numerical method are given in the appendix. 
\begin{figure}[t!]
\includegraphics[width=\linewidth]{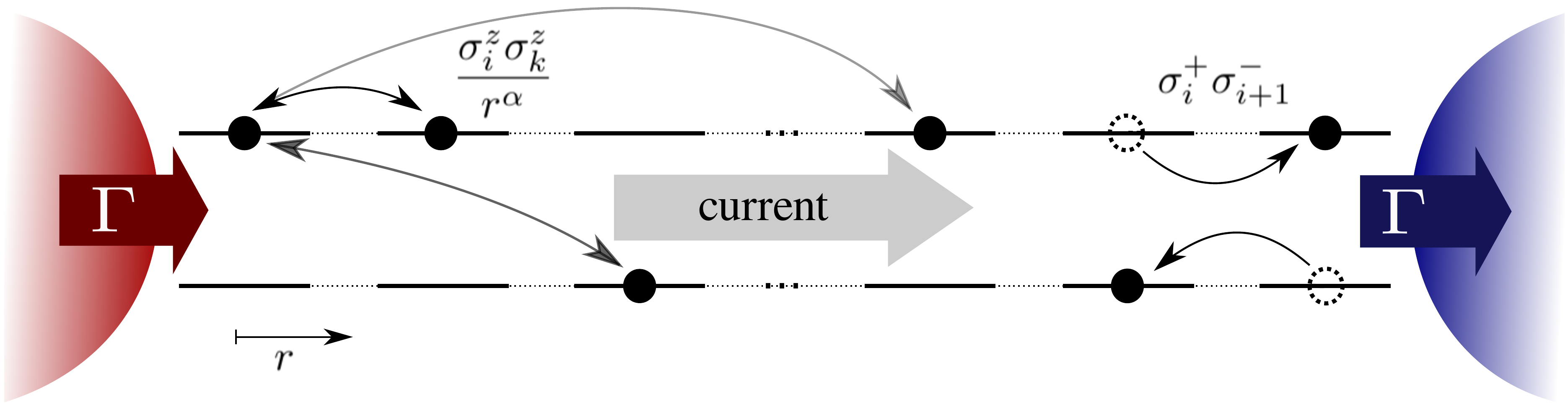}
\caption{Sketch of the spin chain depicted as a chain of TLS, with next neighbour hopping $\sigma_i^-\sigma^+_{i+1}$, and long-range Ising-type interaction $\sigma^z_i\sigma^z_k$, which decreases with distance $r$, mediated by the parameter $\alpha$. The chain is driven by an incoherent Lindblad-type bias at the boundaries with in- and outscattering rate $\Gamma$.}
\label{fig:1}
\end{figure}
\section{Model}\label{modelsection}
This work examines a Heisenberg XXZ chain with nearest neighbour spin-flip interactions as well as long-range Ising-type interaction, a common model in the realm of closed chains \cite{gongprb2016,gongprb2016_2,maghrebiprl2017,vanderstraetenprl,renprb20}. The novelty is to combine this long-range interaction with boundary driving, which was done in a previous work \cite{prbleon17}, albeit more with a focus on robustness and disorder. Here we are interested in the\neu{chain lengths where the transition to ballistic transport for a given long-range interaction strength occurs.}
The Hamiltonian reads
\begin{align}\label{hamilton}
	H = 
	&\sum_{i=0}^{N-2}\frac{J}{4}\big(\sigma_i^x\sigma_{i+1}^x+\sigma_i^y\sigma_{i+1}^y\big)\nonumber\\
	&+\frac{1}{4}\frac{J}{A}\sum_{i=0}^{N-2}\sum_{\ell>i}^{N-1}\frac{1}{|\ell-i|^\alpha}\sigma_i^z\sigma_\ell^z, \\
	A&:=\frac{1}{(N-1)}\sum_{i=0}^{N-2}\sum_{\ell>i}^{N-1}\frac{1}{|\ell-i|^\alpha}
\end{align}
with the weighing constant $A$. 
In our model, the Ising-type interaction occurs not only between neighbouring spins, but also between more distant ones. The interaction strength between site $l$ and $i$ decreases with the distance $r:=|l-i|$ with a power-law parameter $1/r^\alpha$, in order to be able to tune the range, as it is possible in related experiments \cite{Britton2012,Islam2013,Richerme2014,Jurcevic2014,SmithNatPhys2016,Neyenhuise2017,jurcevicprl2017}. The weighing constant $A$ ensures that in total, the interaction strength on each spin stays the same for different chain lengths N and different $\alpha$, and hence the ratio between Ising- and spin-flip-interaction remains the same for the different scenarios. Without this weighing constant, it would not be possible to know whether the transport transition is induced by the long-range or by the anisotropy of the chain. This is important, since an anisotropic ratio between $J_x$ and $J_z$ can also lead to a ballistic transport regime  \cite{znidaricprl11,znidaricprl2016,benentiprb09}. We design the Hamiltonian such that in total, for every site the Ising- and the spin-flip-interaction are equally strong.\neu{This allows us to show that even in this extremely isotropic case, ballistic transport occurs, as soon as the Ising interaction reaches out long enough. We thus explicitly exclude any possible additional contributions from dipole-dipole interaction, or other spin-spin interactions, in order to ensure this isotropy. For any kind of experimental realisation, further effects would possibly have to be considered, dependent on the respective experimental setup.} 

When the long-range parameter is large, e.g. $\alpha=1000$, we reproduce the fully isotropic case, as we apply the same coupling rate $J$ to both parts of the Hamiltonian ($J_x=J_z$). The smaller $\alpha$, the more effective the Ising-type interaction can reach through the chain, leading to a qualitatively different transport behaviour, also in the non-equilibrium steady state (NESS). 
In order to get some intuition, for a value of $\alpha<0.1$, the Ising interaction is about the same between neighbouring sites and those two or three sites apart from each other. This leads to a complete out-cancelling of this interaction, and the chain behaves completely similar to the Heisenberg XX case without Ising, which can be explained by studying the eigensystem of the Hamiltonain (cf.~App.~\ref{currentandeigensystemappendix}).\neu{As soon as we reach $\alpha>10$, the behaviour is no longer distinguishable from the isotropic case, which we simulate by setting the parameter to an arbitrary large $\alpha=1000$.} 
The Heisenberg-Hamiltonian is mathematically similar to a system of spinless fermions (for the transformation see e.g. \cite{benentiprb09}), which is why in Fig.~\ref{fig:1} we depict a chain of two level systems in order to illustrate the different interactions.
We drive the chain incoherently at its boundaries with Lindblad-type baths, and thus the master equation of the system reads
\begin{align}\label{mastereq}
\dot \rho(t)
=
-\frac{i}{\hbar}[H,\rho(t)]
+\frac{\Gamma}{2}\big(\mathcal{D}[\sigma_0^+]\rho(t)
+\mathcal{D}[\sigma_{N-1}^-]\rho(t)\big),
\end{align} 
\neu{with $\rho=\sum_i c_i \ket{\psi_i}\bra{\psi_i}$ and}the Lindblad superoperator $\mathcal{D}[\hat{x}]\rho=2\hat{x}\rho\hat{x}^\dagger-\{\hat{x}^\dagger\hat{x},\rho\}$. Since we are interested in the nonlinear case beyond the very weak driving scenario, we only apply two Lindblad baths, instead of four, as it is often the case in related work. This means that our driving is similar to the case of $f=1$ in \cite{benentiprb09} or $\mu=1$ in \cite{prbleon17}, respectively. We hence expect to always drive the chain at its transport limit, since all excitation which is inserted to the right has to travel through the chain, as the influx is not accompanied by a sink at the same site. This is discussed in more detail in Sec.~\ref{gammasection}.
Note that in this work, Eq.~(\ref{mastereq}) is solved directly only for benchmarking purposes. The actual simulations in the Monte Carlo framework are based on a respective quantum jump simulation. Those turn out to be even more effective in the strong-driving regime than in the weak, in which four instead of two Lindblad-operators become necessary. For details on the method, cf. App.~\ref{qmcappendix}.
Due to the incoherent driving at the boundaries, and the highly symmetric Hamiltonian, after an initial transient behaviour, the system dynamics will reach a NESS with a constant current through the chain from one bath to the other. There has been no numerical evidence that this NESS is not unique, as the Lindblad-superoperator and Hamiltonian connects the full Hilbert space \cite{Nigro_2019}. In all our simulations, the NESS was thus independent of the initial conditions.
This current which, dependent on the experimental setup, could e.g. be transport of magnetization~\cite{znidaricprl11}, will be the main figure of merit for our examinations. It has to be equal between each two sites of the chain, since the sites are all alike and there is no additional loss or gain involved in our setting, which provides a good cross-check for the numerical simulations. 
Using a continuity equation \cite{prbleon17}, we define this spin current as 
\begin{align}\label{currentdefinitionequation}
\langle j \rangle := \frac{J}{2}\langle(\sigma_k^x\sigma_{k+1}^y-\sigma_k^y\sigma_{k+1}^x)\rangle \quad (\forall k \in \{1,..,N-1\}).
\end{align}

In a \textit{ballistic} transport regime, the current is independent on the chain length. If, by contrast, the current does decrease with the chain length, this is usually referred to as \textit{diffusive} transport, and modeled by a power law $\langle j\rangle\propto \frac{1}{N^{-\gamma}}$.\neu{Strictly speaking, one can differentiate between \textit{sub-} ($\gamma>1$) and \textit{superdiffusive} ($\gamma<1$) transport, however in the following we refer to all of these regimes as \textit{diffusive} transport.
The NESS current in our case proved to be completely independent of the initial conditions,}and it proved effective to initialize the simulation not in an alldown-state, but with the first site initially in the spin-up state
\begin{align}
    \ket{\psi(t=0)} = \ket{\uparrow\downarrow\downarrow...}.
\end{align}

All other tested initial states showed the same results in the NESS, but longer transients. Unlike in other numerical methods, in our case it also proved ineffective to start the dynamics close to the steady state value. This is due to the fact that before the averaging, the individual Monte Carlo trajectories are not even close to the steady state value, cf. App.~\ref{qmcappendix}.
%

\section{Results}\label{centralresultssection}
\subsection{Transition from diffusive to ballistic transport}
\begin{figure}[t!]
	\includegraphics[width=\linewidth]{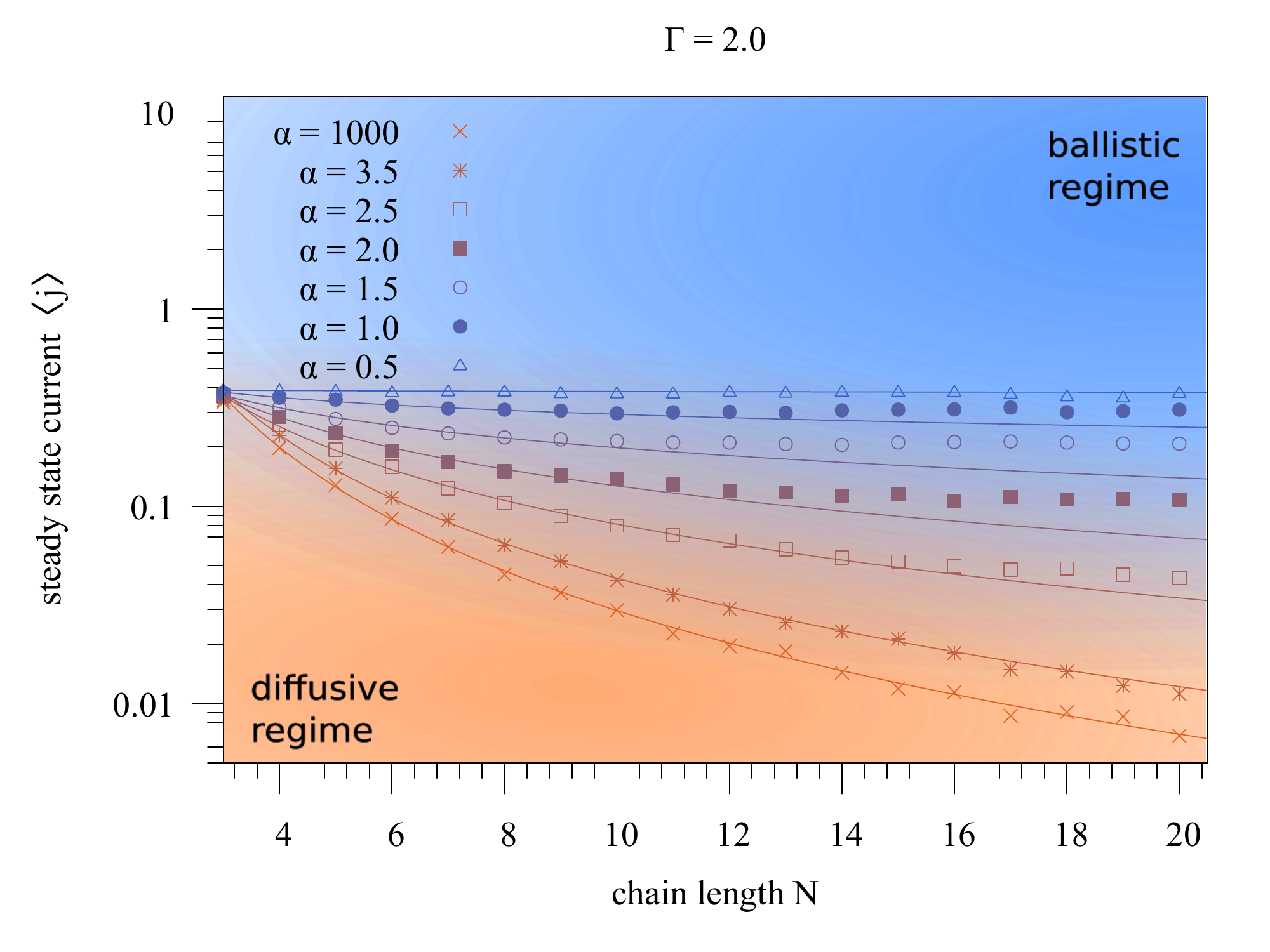}
	\caption{Logplot of the steady state values of the spin current $\langle j\rangle$ over the chain length N for different values of the long-range-parameter $\alpha$. In the diffusive regime, the data is fitted with a power-law fit $1/N^{-\gamma}$. For small $\alpha$, the current enters the ballistic regime already for small $N$, i.e. the current no longer changes for increasing chain lengths. With larger values of $\alpha$, i.e. less long-range interaction, the\neu{saturation occurs only for significantly longer chain lengths.}}\label{fig:2}
\end{figure}
A Heisenberg XX chain without Ising-interaction shows ballistic transport behaviour for a Lindblad-driving scenario \cite{ZnidaricJPA2010,znidaricprl11}. Thus the spin flip interaction shows the same NESS current independent of the chain length. In the case of a Heisenberg XXZ chain, the Ising interaction hinders the transport, and hence, for the same driving, the NESS current decreases with the chain length, which is called diffusive (or in the isotropic case superdiffusive \cite{znidaricprl11,znidaricprl2016}) behaviour. 
While the\neu{isotropic nearest-neighbour chain}does not show ballistic transport even in the thermodynamic limit of very long chains, anisotropic chains do saturate to a ballistic regime after certain chain lengths \cite{znidaricprl11,prosenprl11}.\neu{Like in these works, the transition we observe here does not show the characteristics of a sharp phase transition, but rather manifests as a smooth saturation towards a constant value of the current for long enough chains. We want to stress that this transition occurs in a far-from equilibrium regime and does not show characteristics of a dissipative phase transition \cite{kessleretal2012DPT} or the like.}
It was shown in Ref.~\cite{prbleon17} that long-range Ising interaction changes the transport behaviour from diffusive to ballistic transport.
For strong long-range interaction (e.g. $\alpha=0.1$), the chain length does not have an effect on the steady state current, i.e. the system shows a ballistic transport behaviour for all chain lengths. This means that for extreme long-range, the Ising interaction no longer hinders the spin-flip interaction in its ballistic transport behaviour. 
We show here that for intermediate values of $\alpha$, we can identify a transition point between those two regimes. 
\begin{figure}[t!]
	\includegraphics[width=\linewidth]{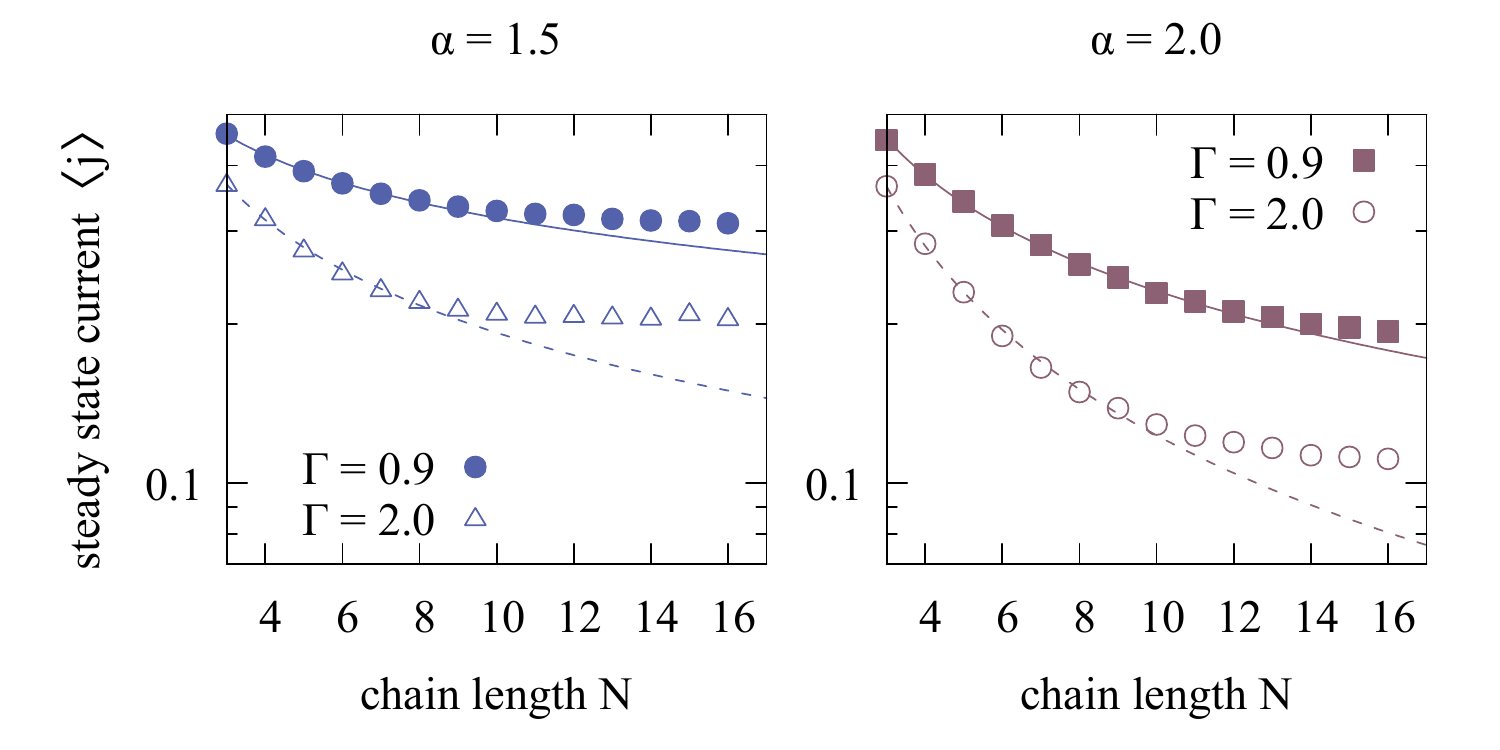}
	\caption{Logplot of the steady state values of the relative current j over the chain length N for different driving strenghts (scattering rate $\Gamma$), the two plots show different long-range scenarios $\alpha$. \neu{The chain length dependence for the transition to}ballistic transport is much more affected by $\alpha$ than by $\Gamma$.}\label{fig:3}
\end{figure}

Dependent on the strength of the long-range interaction,\neu{we can identify chain lengths}where the transport changes from diffusive to ballistic behaviour, thus after a certain chain length, the current does not decrease any more.
\neu{These chain lengths are}directly dependent on $\alpha$, hence on the range of the long-range Ising interaction, and relatively independent of the in- and outscattering rate $\Gamma$.
Fig.~\ref{fig:2} shows the steady state values of the NESS current $\langle j\rangle$ over N for different long-range parameters $\alpha$. In this plot, all values are obtained for an in- and outscattering rate of $\Gamma=2.0$, and throughout the paper we set $\hbar = 1$ and $J=1$. In the diffusive regime, the data points are fitted with a power-law fit $1/N^{-\gamma}$, following the definition of diffusive transport. At the transition point, the steady state values no longer follow this fit, but instead remain at the same level when N is further increased, i.e. the system becomes ballistic. As $\alpha$ grows (which means decreasing long-range interaction), the\neu{range}of N where this occurs is shifted towards longer chain lengths. Note that a similar transport transition for growing chain lengths was found also for anisotropic chains \cite{znidaricprl11}. We want to stress that due to our weighing constant A in Eq.~(\ref{hamilton}), this chain is not anisotropic in a sense of an imbalance between spin-flip and Ising-type interaction. Thus, the effect here must be due to the longer range of the Ising interaction, beyond the nearest neighbour case. This is also supported by the findings on the effect of long-range Ising interaction on the eigenvectors of the Hamiltonian, cf.~Sec.~\ref{discussionsection} and App.~\ref{currentandeigensystemappendix}. 
One could expect that the \neu{transition chain lengths are}highly dependent on the driving strength at the boundaries, mediated by the in-/outscattering $\Gamma$, especially since other anomalous transport effects such as negative differential conductivity (NDC)  strongly depend on the driving \cite{benentiprb09,prbleon17}. However, we find that the transition to ballistic transport, although highly dependent on $\alpha$, takes place more or less\neu{at the same chain lengths}for all driving scenarios (at least above the linear regime of very weak driving).
%

\subsection{Independence of the driving strength $\Gamma$}\label{gammasection}

\begin{figure}[b!]
	\includegraphics[width=\linewidth]{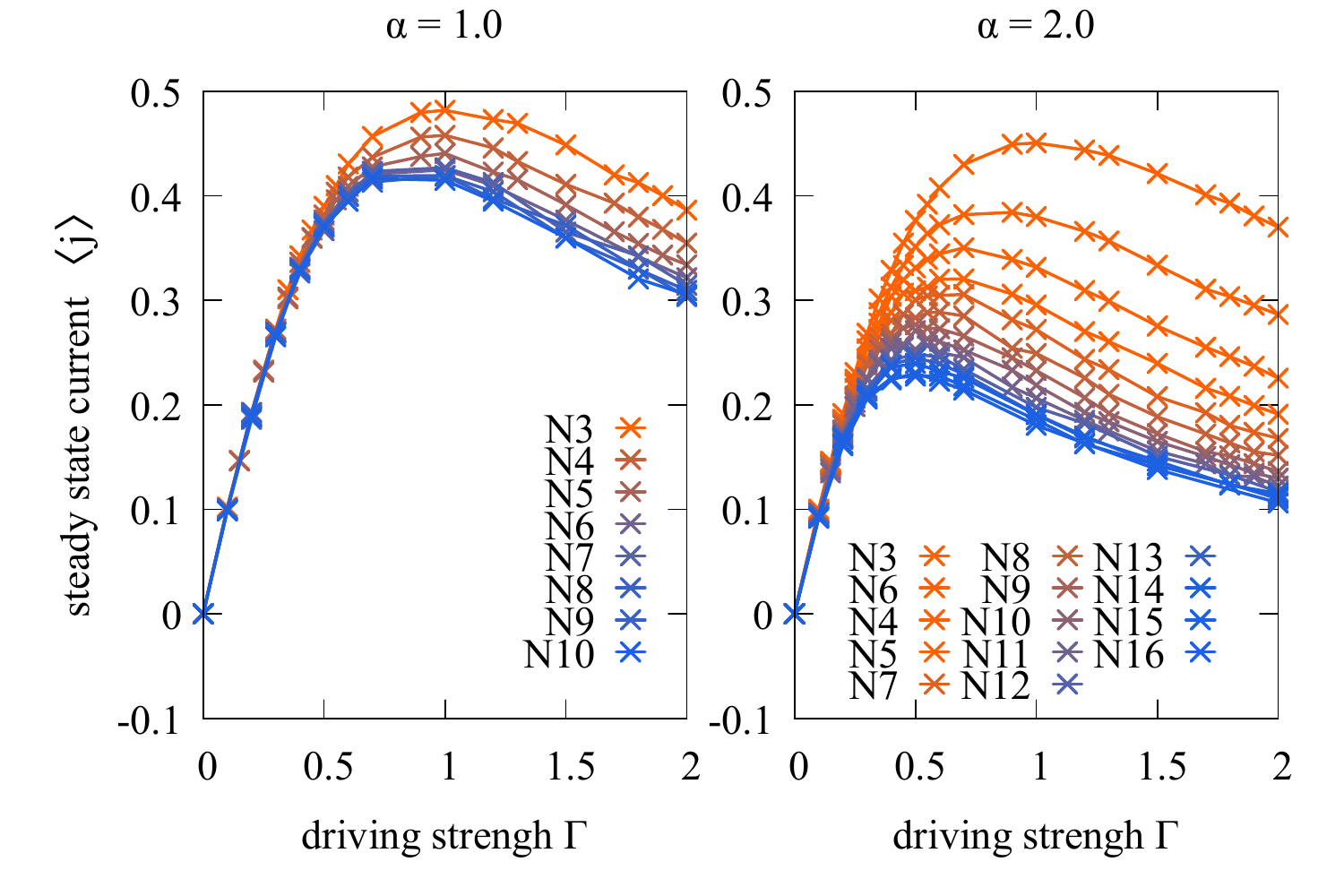}
	\caption{Steady state values of the spin current $\langle j\rangle$ over the in/outscattering rate $\Gamma$ of the incoherent driving for different chain lengths N. The two plots show scenarios of different long-range parameters $\alpha$. Ballistic transport is reached, when the curves for different N are no longer distinguishable, for $\alpha = 2.0$ around\neu{$N\approx15$,}for $\alpha=1.0$ already for\neu{$N<10$.}We show here that as soon as $\Gamma$ leaves the linear regime of very weak driving, the transport transition point no longer qualitatively depends on the driving strength.}
	\label{fig:4}
\end{figure}
While a larger $\alpha$ drastically shifts the transport transition point towards longer chains, the scattering rate $\Gamma$ of the incoherent driving bias has a rather small influence on this phenomenon, as will be shown in this section. Fig.~\ref{fig:3} shows the transition towards the ballistic transport regime for two different inscattering-rates. The direct comparison demonstrates that the driving changes the absolute value of the current in the chain without changing \neu{the chain lengths} for the transition to the ballistic regime. 
Note that this paper is focused on far-from-equilibrium effects above the linear regime of very weak driving. 
The driving is thus considered strong enough to show negative differential conductivity (NDC). In this driving regime, the transport transition of the NESS current is not qualitatively dependent on the driving $\Gamma$ since the driving is always strong enough to be able to identify the current inhibiting bottleneck in the chain. 
\begin{figure}[t!]
\includegraphics[width=\linewidth]{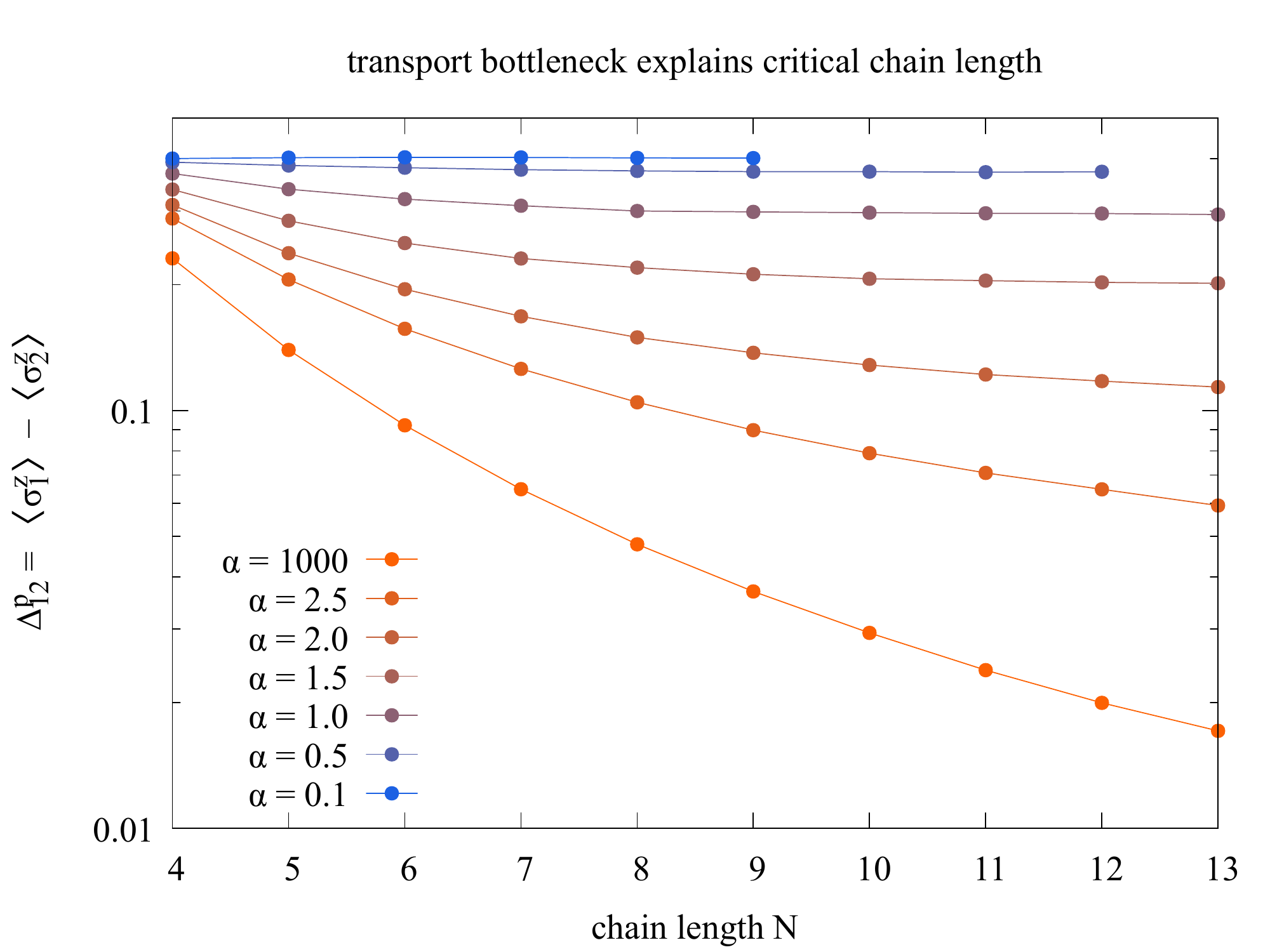}
\caption{The difference between the polarisations of the first and second chain site mark the bottleneck for the transport. This value is plotted here over N for different long-range scenarios $\alpha$, in a logplot to stress its similarity to Fig.~\ref{fig:2}. When this bottleneck no longer changes for growing chain lengths, the system enters the ballistic regime. The transition is shifted to longer chains for larger $\alpha$, which fits the findings in Fig.~\ref{fig:2}.}
\label{fig:5}
\end{figure}
This is underlined by Fig.~\ref{fig:4}, which shows the steady state current over the driving $\Gamma$ for two different intermediate long-range scenarios. The first plot shows $\alpha=1.0$, where the interaction scales reciprocal with the distance. We find that already for chain lengths $N<10$ the transport is ballistic. In the case of $\alpha=2.0$, the ballistic regime is only reached for a chain length of\neu{$N\approx15$.}The plots show that this occurs for all driving scenarios $\Gamma$ around the same\neu{chain lengths}(at least beyond the linear regime of very small driving, where the current is very small and the chain thus shows ballistic transport already for very short chains).
For both plots of Fig.~\ref{fig:4}, there is an inscattering rate $\Gamma_{max}$ which yields a maximum current, and for stronger driving the current decreases again. This effect is called negative differential conducticity (NDC), which occurs due to current inhibiting effects \cite{benentiprb09,prbleon17}. We are interested in a regime, where the driving is strong enough to make those effects visible. We now discuss in detail the transition from diffusive to ballistic transport for such regimes.
\section{Discussion}\label{discussionsection}
Here we attempt a physical explanation for the result we obtained from our numerical studies, namely the transition from diffusive to ballistic transport, and its strong dependence on $\alpha$. The section is organized as follows: First we present the difference between the Heisenberg XX and the Heisenberg XXZ chain in a boundary-driven scenario, and show why only with Ising interaction we see non-ballistic transport, due to the formation of ferromagnetic domains. We also give reasons, why the transport bottleneck will always be found at the edges of the chain. In the second part, we argue why long-range Ising interaction, in contrast to the isotropic nearest-neighbour case, suppresses those domains, and hence the system can become ballistic again. The third part finally gives arguments for the strong $\alpha$-dependence of\neu{the chain length}where this transition occurs. We introduce the bottleneck polarisation difference $\Delta^p_{12}$   between the first and the second site of the chain, cf.~Fig.~\ref{fig:5}, which saturates at\neu{the transition chain length,}dependent on $\alpha$, just as it is the case for the steady state current, and thus explaining the transition to ballistic transport.  
\subsection{Ferromagnetic domains hinder ballistic transport}
\begin{figure}[t!]
	\includegraphics[width=.99\linewidth]{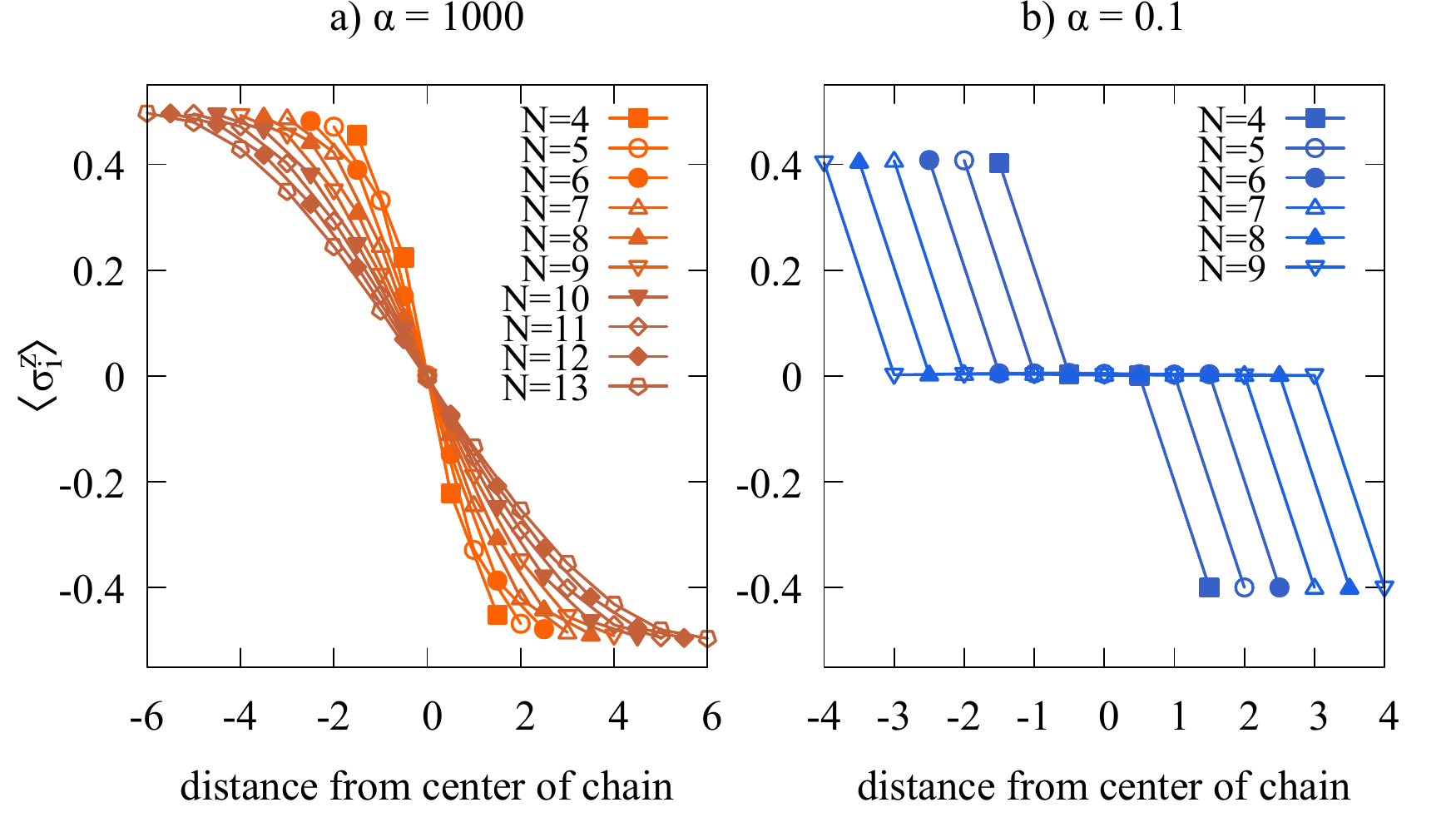}
	\caption{a) Polarization profiles for different chain lengths in the isotropic case without long-range, $\alpha=1000$. Due to the nonzero Ising interaction, the sites at the edges of the chain are quite well aligned, and align better and better with growing chain lengths. This alignment hinders the current, which explains the non-ballistic behaviour even in the thermodynamic limit of very long chains \cite{znidaricprl11}. b) Opposite case of extreme long-range, $\alpha=0.1$, the polarisation profile exactly resembles the Heisenberg XX chain \cite{bertiniARXIV2020}, with nonzero polarisation only at the edges of the chain and thus fully ballistic transport.}
	\label{fig:6}
\end{figure}
Analyzing the definition of the spin current in Eq.~\eqref{currentdefinitionequation}, one finds that the current is only realized by a certain part of the basis states, while others cannot support a transport through the chain, namely those where the spins are both aligned in the same direction and parallel to the $z$-axis. 
As a consequence, the bigger the difference between the polarisation of the first two (or the last two) spins in the chain (cf.~App.~\ref{currentandeigensystemappendix}), the larger the current. 
Hence, a Heisenberg XX chain without Ising interaction shows completely ballistic transport properties in a boundary driving scenario~\cite{znidaricprl11}\neu{(which can even be verified by an analytic solution via a Matrix Product Operator ansatz~\cite{ZnidaricJPA2010}). Its}polarisation profile in the steady-state shows edge spins which are strongly polarised in the direction of the driving baths ($\langle\sigma_1\rangle\gg0\gg\langle\sigma_N\rangle$) while all other spins are polarised perfectly perpendicular to the z-axis ($\langle\sigma_i^z\rangle=0,\quad 1 < i < N$). 
The polarisation profiles then look alike for all chain lengths $N>2$, as only the edge spins have a special role. This fits to the numerical observation that the current does also not change when prolonging the chain, hence the system is completely ballistic.
The Heisenberg XX case is important to understand what happens in a XXZ chain with long-range, as we will see in the following. 
In the Heisenberg XXZ chain, where Ising interaction is present, for $N=2$, the polarisation or steady-state current is not changed in comparison to the XX chain (cf.~corresponding eigenvectors in App.~\ref{currentandeigensystemappendix}).
However, for Heisenberg XXZ chains with $N\geq 3$, a nonzero nearest neighbour Ising interaction strongly inhibits the current due to the NDC \cite{znidaricprl11,prbleon17}. This is unraveled by the eigensystem of this chain: A Hamiltonian with Ising nearest neighbour interaction has different eigenvectors compared to the case of only spin-flip interaction  cf.~App.~\ref{currentandeigensystemappendix}. It is also well-known \cite{benentiprb09,prbleon17} that for longer chains, the polarisation profiles of the XXZ chain show ferromagnetic domains at the edges of the chain, cf.~Fig.~\ref{fig:6}a. Those domains are mostly dominated by the $\ket{\uparrow\uparrow..}$ or $\ket{\downarrow\downarrow..}$
eigenvectors of the Hamiltonian, which by definition of the current do not contribute to the energy transport. This explains why those domains are hindering the transport. 
Fig.~\ref{fig:6}a shows that the longer the chain, the stronger becomes the alignment between the edge spins and therefore the NESS current is suppressed for growing chain lengths (diffusive behaviour). In the\neu{case of a fully isotropic chain ($J_x=J_z$) with an Ising interaction only reaching to the nearest neighbour,}the current stays non-ballistic even in the thermodynamic limit of very long spin chains~\cite{znidaricprl11,znidaricprl2016}. \neu{(Analytic calculations also suggest that the isotropic chain 
cannot become ballistic, not even for infinitely long chains~\cite{prosenprl11}.)}However, with long-range Ising interaction, ballistic transport can be recovered.
%

\subsection{Long-range suppresses the ferromagnetic domains}\label{polarisationsection}
\begin{figure}[t!]
\includegraphics[width=\linewidth]{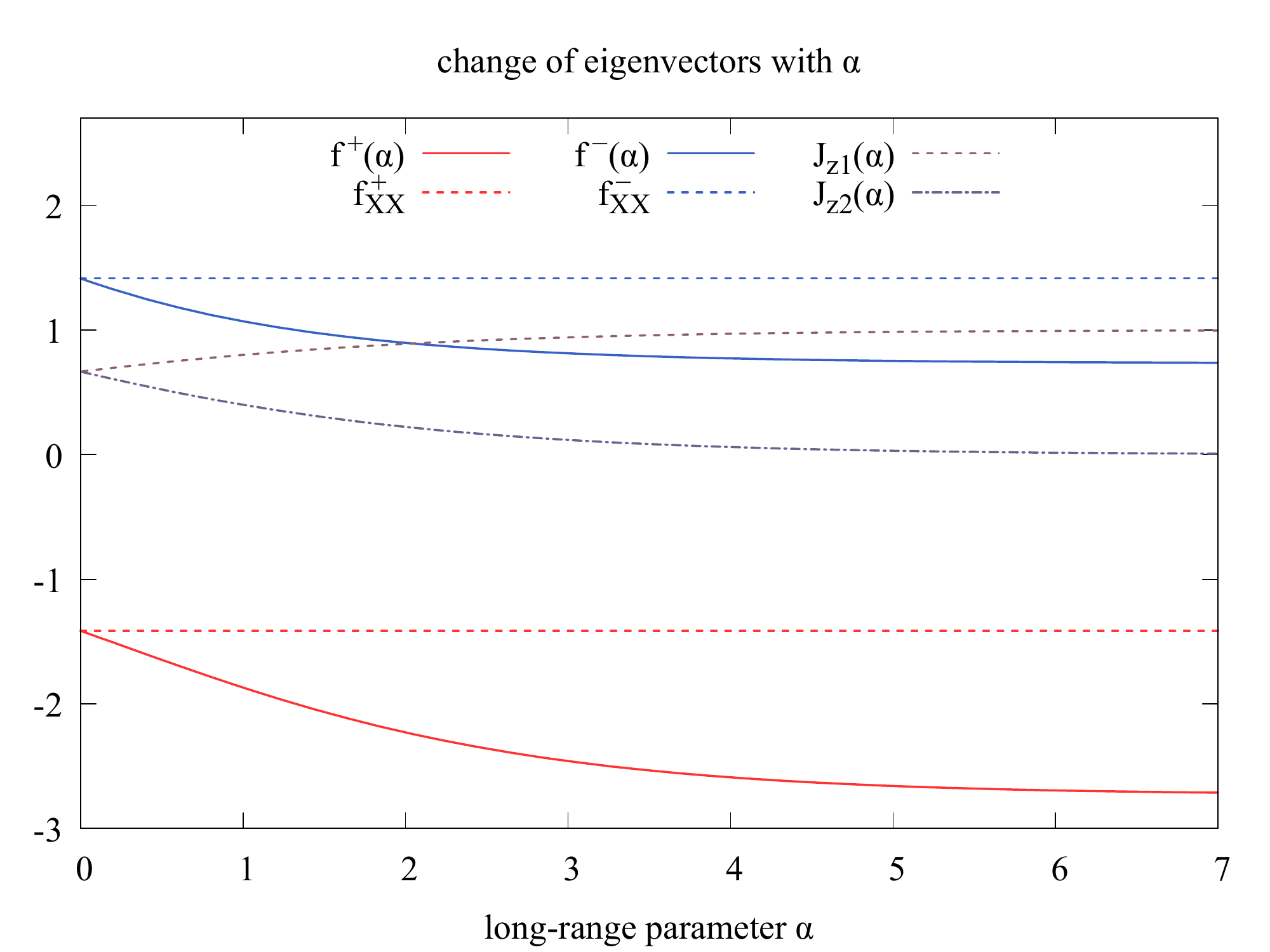}
\caption{Change of eigenvectors with the long-range parameter $\alpha$ for the Hamiltonian of a chain with N=3, details cf.~App.~\ref{currentandeigensystemappendix}. The plot shows also the case without Ising interaction ($f^\pm_{XX}$), which is reproduced for the long-range XXZ chain for very small $\alpha$, since the Ising interaction cancels itself out. (Note that this effect can be observed for all ratios of $J_z/J_x$, but is more drastic for more anisotropy. For this plot, we set $J_x=0.5$.) We also show the the $\alpha$-dependence of $J_{z1,2}$.}
\label{fig:7}
\end{figure}

In contrast to the nearest-neighbour case, for long-range interaction,\neu{after a certain chain length,}we observe a transition to the ballistic transport regime. In an extreme long-range scenario ($\alpha=0.1$), where effectively, the Ising interaction is the same between neighbouring sites and the next ones following further away in the chain, we numerically reproduce the Heisenberg XX case, with ballistic transport already for chains of smallest lengths, and a polarisation profile with only the first and last spin showing a nonzero $\langle \sigma^z \rangle$ polarisation, cf.~Fig.~\ref{fig:6}b.
The reason is that the Ising interaction out-cancels itself and no longer hinders the ballistic transport mediated by the spin-flip interaction. This is also supported by the fact that for extremely small $\alpha$, the Hamiltonian for a chain of $N=3$ shows exactly the same eigenvectors as in the case without Ising interaction, cf.~Fig.~\ref{fig:7} and App.~\ref{currentandeigensystemappendix}.
While for extreme cases of $\alpha<0.1$, this already occurs for smallest chain lengths, for an intermediate range of the Ising interaction, ballistic behaviour is only recovered for longer chains, which will be discussed in the next paragraph.
%

\subsection{Intermediate interaction-range and the transport transition}
\begin{figure}[t!]
	\includegraphics[width=\linewidth]{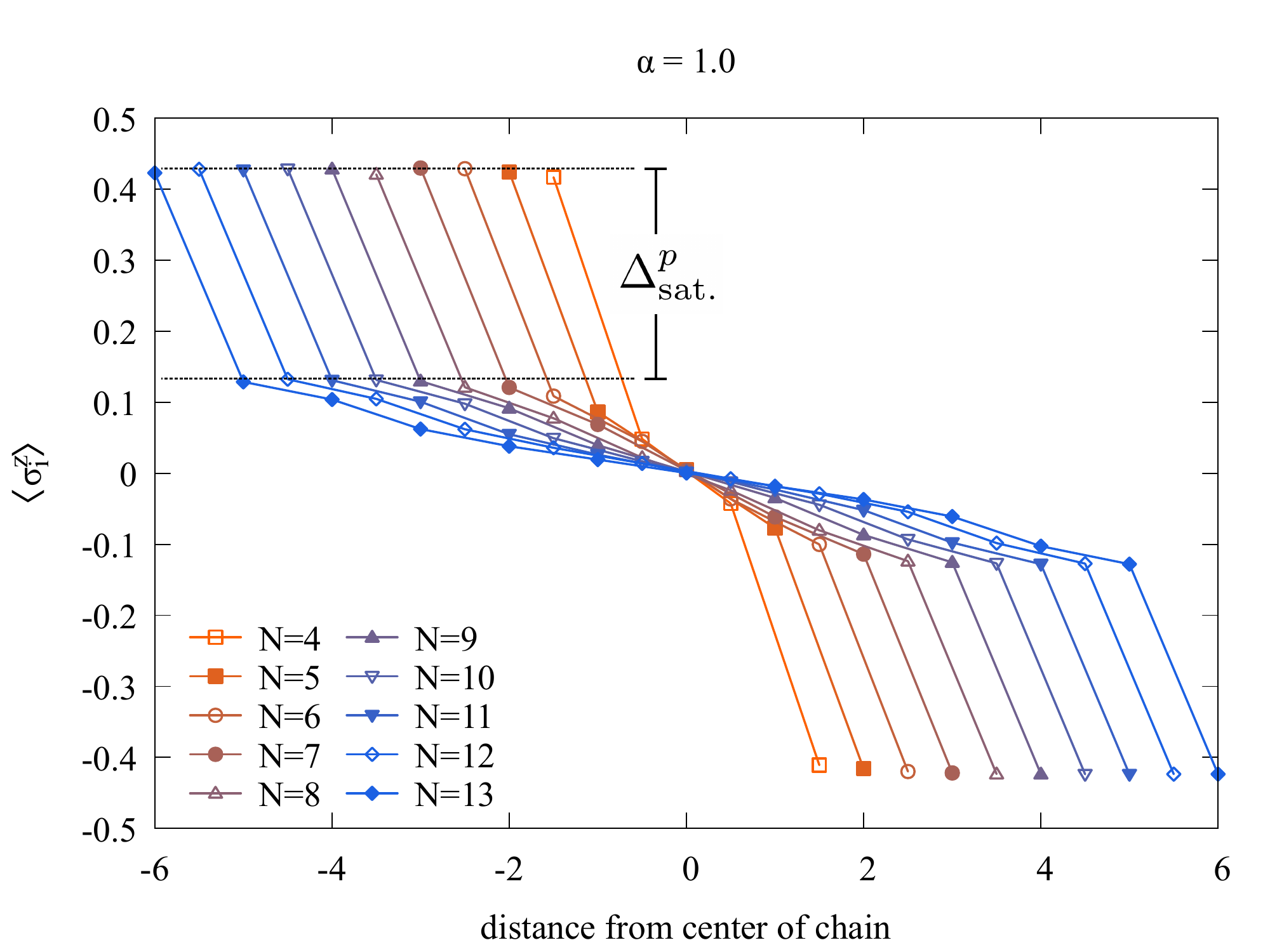}
\caption{Polarisation profiles for $\alpha=1.0$ for chains of different lengths. The less far reaching the interaction, the longer the chain has to be until an increasing of N no longer changes the current defining polarisation difference between the edge sites and their direct neighbours,\neu{and hence the bottleneck value saturates to a constant value $\Delta^p_{\text{sat.}}$.}Here, this point is reached for $N\geq9$.}
	\label{fig:8}
\end{figure}

Now we will focus on intermediate values of $\alpha$, where the interaction has a shorter range, and the nearest neighbour Ising interaction is significantly stronger than the next-nearest neighbour interaction. For those intermediate long-range scenarios, we find a transition to ballistic transport only\neu{once the chain is long enough.}Fig.~\ref{fig:8} shows profiles for long-range interaction reciprocal to the distance ($\alpha=1.0$). The difference of the polarisation between the edge sites and their direct neighbours changes for growing chain lengths, until a certain point, e.g. for $\alpha=0.5$ around\neu{$N\approx5$,}for $\alpha=1.0$ around\neu{$N\approx9$,}and for $\alpha=1.5$ around\neu{$N\approx12$}. After this point, it remains at a constant value (for more plots refer to App.~\ref{polarisationsappendix}).  Since this difference between the polarisation of the edge sites is the bottleneck for the NESS current, it is an indicator for the transition point towards the ballistic transport regime. 
In order to further quantify this, we calculate the difference between the polarisations of the first two sites, a value strongly dependent on $\alpha$ and $N$,  which we showed is the bottleneck of the current. (Note that one could just as well examine the last two sites in the chain, due to the symmetry of the driving and the interactions.) We hence define the bottleneck-difference
\begin{align}\label{bottleneckeq}
    \Delta^p_{12}= 
    \langle \sigma_1^z\rangle -
    \langle \sigma_2^z\rangle. 
\end{align}{}
This provides us a figure of merit allowing us to explain the transition to ballistic transport. It is plotted over N for different $\alpha$ in Fig.~\ref{fig:5}. Note that small values of this bottleneck-figure denote a strong alignment of the first two spins. The first spin will always be strongly aligned due to the driving bath, at least beyond the linear regime of very weak driving. Small values therefore resemble a dominant $\ket{\uparrow\uparrow..}$ - eigenvector of the respective Hamiltonian, and hence a strong hurdle for transport, while for large values we are closer to the Heisenberg XX case which allows the strongest transport.
Fig.~\ref{fig:5} shows the dependence of this bottleneck $\Delta^p_{12}$ on $N$, for different long-range scenarios $\alpha$. It is in very good accordance with the transport regimes we identified in Fig.~\ref{fig:2}. The bottleneck decreases with $N$, until in the ballistic regime, it does not change when further prolonging the chain.\neu{The chain length}where this occurs has the same strong dependence on the long-range parameter $\alpha$, as we found it for the NESS current. Note that for different driving strengths $\Gamma$, the polarisation profiles do not qualitatively change (Fig.~\ref{fig:A2} in App.~\ref{polarisationsappendix}), which fits to the discussed observation that the influence of the driving $\Gamma$ on the transition is small (at least above the linear regime of very weak driving).
Our analysis has shown that nearest-neighbour Ising interaction hinders ballistic transport, as it alters the eigensystem of the spin-flip Hamiltonian and leads to a formation of ferromagnetic domains at the edges of the chain. These domains suppress the transport through the chain. The edge spins align more and more for longer chain lengths and diffusive behaviour is inevitably the consequence. In contrast, for extreme long-range, the eigensystem of the ballistic Heisenberg XX chain without Ising interaction is recovered, and also the polarisation profile resembles the case without Ising interaction. For intermediate long-range, the bottleneck at the edges of the chain changes for small chains, but saturates\neu{at a $\alpha$-dependent chain length,}which explains why the transition to ballistic transport occurs at those chain lengths.
%

\section{CONCLUSION AND OUTLOOK}
We showed that when boundary-driving a Heisenberg XXZ chain with nearest-neighbour spin flips, and an Ising-type interaction exceeding the nearest-neighbour case, a ballistic transport regime is reached after a\neu{certain chain length,}where the transport no longer decreases.\neu{This chain length}is smaller for longer-ranging Ising interaction, and this effect is independent of the driving strength $\Gamma$.
The reason for the transition can be readily explained by the polarisation profiles of the chain. The long-range interaction suppresses the ferromagnetic domains and hence enlarges the bottleneck for the NESS current at the edges of the chain.\neu{Once the chain is long enough, this bottleneck saturates and}the transition towards ballistic transport is reached.
\neu{Neither our numerics, nor the presented analysis give any indication why the system should not stay ballistic for even longer chains, or even in the thermodynamic limit. However, a numerical simulation of longer chains would be very much appreciated to verify this, maybe e.g. with the use of the upcoming neural network techniques which are currently developed for simulations of open quantum systems~\cite{savona2019NN,Hartmann2019NN,Cuiti2019NN}. Besides, examinations on finite temperature effects~\cite{Prosen2009temperature,jesenko2011temperature} and disorder~\cite{prbleon17,znidaricprl2016} would be very interesting to apply to the model we presented here, with the latter arising intriguing questions about many body localisation~\cite{nandkishorehuse2015mblreview}.}
\section*{Acknowledgements}
We gratefully acknowledge support of the Deutsche Forschungsgemeinschaft (DFG) through project B1 of the SFB 910,
and we thank Markus Heyl and Andreas Knorr for stimulating discussions in the first stages of the project.
\appendix
\section{The relation of eigensystem and spin current}\label{currentandeigensystemappendix}
Following \cite{Parkinson2010}, we introduce a Hilbert space for two spins:
\begin{align}
    \ket{\psi}=
    c_1\ket{\uparrow\uparrow}
    +c_2\ket{\downarrow\downarrow}
    +c_3\ket{\uparrow\downarrow}
    +c_4\ket{\downarrow\uparrow}
\end{align}{}
The current is defined as stated in Eq.~(\ref{currentdefinitionequation}) in the main text. When applying the above basis we end up with
\begin{align}
    \bra{\psi}j\ket{\psi}
    \propto
    c_3^*c_4-c_4^*c_3.
\end{align}{}
This means that only the last two basis vectors contribute to the current, and both of them need to have finite probabilities for a nonzero spin current. The ferromagnetic domains mentioned in Sec.~\ref{discussionsection} of the main text show two or more neighbouring spins which are strongly polarized in the z-direction, and will hence be dominated by either $\ket{\uparrow\uparrow}$ or $\ket{\downarrow\downarrow}$, which can both not contribute to the current. This explains the current inhibiting effect of those ferromagnetic domains and hence the diffusive behaviour of the nearest neighbour XXZ chain.\\
\\
The Hamiltonian
\begin{align}
    H
    =\frac{J_x}{2}(\sigma_1^+\sigma_2^-+h.c.)+J_z\sigma_1^z\sigma_2^z.
\end{align}{}
in this basis has the following eigenvectors
\begin{align}
    &\ket{\psi_1}=\ket{\uparrow\uparrow}\qquad
    \ket{\psi_2}=\ket{\downarrow\downarrow}\nonumber\\
    &\ket{\psi_3}=\ket{\uparrow\downarrow}+\ket{\downarrow\uparrow}\nonumber\\
    &\ket{\psi_4}=\ket{\uparrow\downarrow}-\ket{\downarrow\uparrow}.
\end{align}{}
Note that for $N=2$, these eigenvectors occur for all relations of $J_x$ and $J_z$, and even for $J_z=0$. From the numerics we see that in the NESS, neither the polarisation nor the current are influenced by $J_z$ when we only have two spins.\\
\\
For three or more spins we see in the numerics, that Ising nearest neighbour interaction has a current inhibiting effect, and changes the polarisation profile compared to the XX case without Ising interaction. We hence also analyse the eigensystem of the Hamiltonian of the $N=3$ chain. For this we need a larger Hilbert space
\begin{align}
\ket{\psi}
=&\quad c_1\ket{\uparrow\uparrow\uparrow}
+c_2\ket{\uparrow\uparrow\downarrow}
+c_3\ket{\uparrow\downarrow\uparrow}
+c_4\ket{\uparrow\downarrow\downarrow}\nonumber\\
&+c_5\ket{\downarrow\uparrow\uparrow}
+c_6\ket{\downarrow\uparrow\downarrow}
+c_7\ket{\downarrow\downarrow\uparrow}
+c_8\ket{\downarrow\downarrow\downarrow}.
\end{align}{}
The respective Hamiltonian reads
\begin{align}
    H=&\frac{J_x}{2}(\sigma_1^+\sigma_2^- + \sigma_2^+\sigma_3^- + h.c.)\nonumber\\ &J_{z1}(\sigma_1^z\sigma_2^z + \sigma_2^z\sigma_3^z)+J_{z2}\sigma_1^z\sigma_3^z.
\end{align}{}
Note that $J_{z1}$ refers to nearest neighbour interaction, while $J_{z2}$ only occurs in the long-range case, i.e. for small values of $\alpha$.\\

We find that  here the XX case of $J_{z1}=J_{z2}=0$ (which shows the same current as for $N=2$) has those eigenvectors
\begin{align}
    &\ket{\psi_{1,2}}=\ket{\uparrow\uparrow\downarrow}\pm\sqrt{2}\ket{\uparrow\downarrow\uparrow}\nonumber+\ket{\downarrow\uparrow\uparrow}\\
    &\ket{\psi_{3,4}}=\ket{\uparrow\downarrow\downarrow}\pm\sqrt{2}\ket{\downarrow\uparrow\downarrow}+\ket{\downarrow\downarrow\uparrow}\nonumber\\
    &\ket{\psi_{5}}=\ket{\uparrow\downarrow\downarrow}-\ket{\downarrow\downarrow\uparrow}\nonumber\\
    &\ket{\psi_{6}}=\ket{\uparrow\uparrow\downarrow}-\ket{\downarrow\uparrow\uparrow}\nonumber\\
    &\ket{\psi_{7}}=\ket{\uparrow\uparrow\uparrow}\nonumber\\
    &\ket{\psi_{8}}=\ket{\downarrow\downarrow\downarrow}.
\end{align}{}
This is the eigensystem which numerically shows the same NESS current as for $N=2$. In contrast to the $N=2$ case, if we introduce Ising interaction, the current decreases. Interestingly, we find that the first four of the eight eigenvectors also change for nonzero values of $J_{z1,2}$:
\begin{align}
    &\ket{\psi_{1,2}}=\ket{\uparrow\uparrow\downarrow}-f^\pm(J_x,J_{z1},J_{z2})\ket{\uparrow\downarrow\uparrow}\nonumber+\ket{\downarrow\uparrow\uparrow}\\
    &\ket{\psi_{3,4}}=\ket{\uparrow\downarrow\downarrow}-f^\pm(J_x,J_{z1},J_{z2})\ket{\downarrow\uparrow\downarrow}+\ket{\downarrow\downarrow\uparrow}\nonumber\\
    &\ket{\psi_{5}}=\ket{\uparrow\downarrow\downarrow}-\ket{\downarrow\downarrow\uparrow}\nonumber\\
    &\ket{\psi_{6}}=\ket{\uparrow\uparrow\downarrow}-\ket{\downarrow\uparrow\uparrow}\nonumber\\
    &\ket{\psi_{7}}=\ket{\uparrow\uparrow\uparrow}\nonumber\\
    &\ket{\psi_{8}}=\ket{\downarrow\downarrow\downarrow}
\end{align}{}
with 
\begin{align}
    &f^\pm(J_x,J_{z1},J_{z2})\nonumber\\
    &=
    \frac{J_{z1}-J_{z2}\pm\sqrt{8J_{x}^2+J_{z1}^2-2J_{z1}J_{z2}+J_{z2}^2}}{2J_x}.
\end{align}{}
Now for the extreme case of $\alpha<0.1$, i.e. a very long-ranging Ising interaction and thus approximately $J_{z1}\approx J_{z2}$, we find that 
\begin{align}
    f^\pm(J_x,J_{z1},J_{z2})\approx\pm\sqrt{2}
\end{align}{}
and hence we reproduce the eigenvectors of the XX case. Ising thus outcancels itself, which explains why the current is no longer inhibited, and the ballistic behaviour is restored. Fig.~\ref{fig:6} of the main text shows how for small $\alpha$, the values of $J_{z1}$ and $J_{z2}$ become more and more similar, and how this leads to an outcancelling of the Ising interaction, as the eigenvectors of the Heisenberg XX case are being restored.
\section{Polarisation profiles}\label{polarisationsappendix}
\begin{figure}[h!]
	\includegraphics[width=\linewidth]{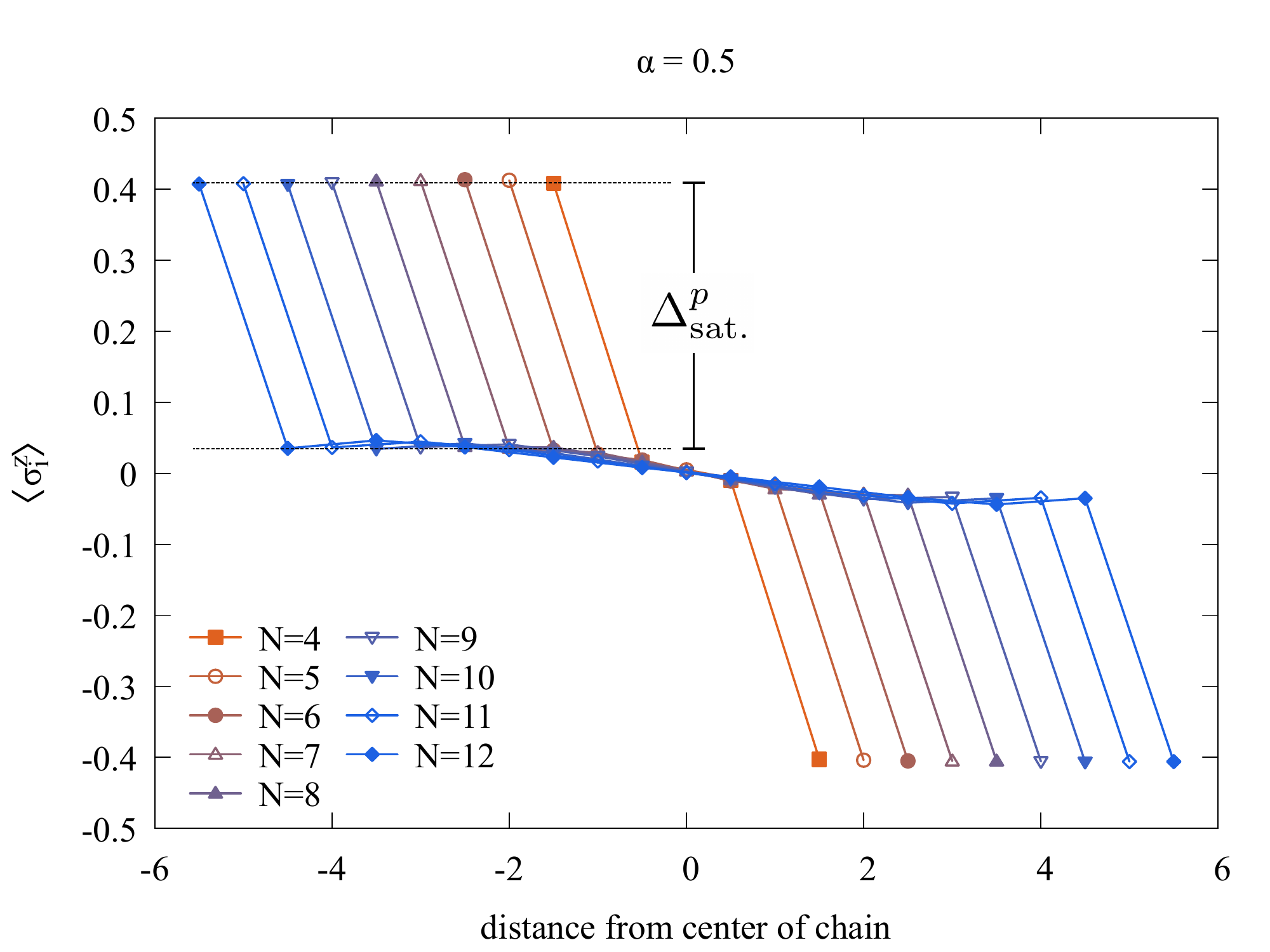}
	\includegraphics[width=\linewidth]{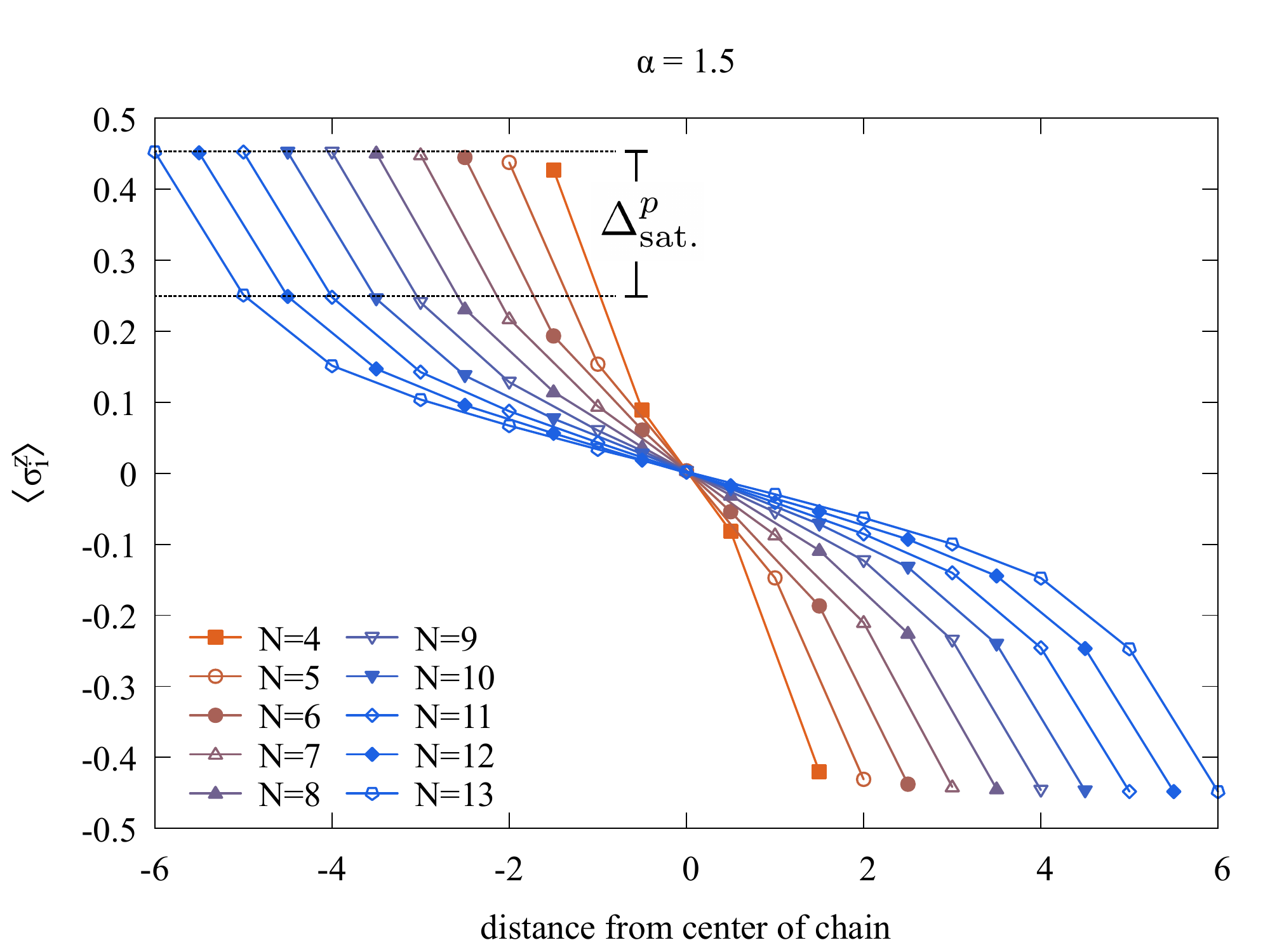}\caption{Polarisation profiles for different chain lengths, the two plots show different long-range scenarios. The \neu{saturation} bottleneck-polarisation difference at the edges is reached\neu{at the transition chain length,}which directly depends on $\alpha$.}
	\label{fig:A1}
\end{figure}
\begin{figure}[htbp]
	\includegraphics[width=\linewidth]{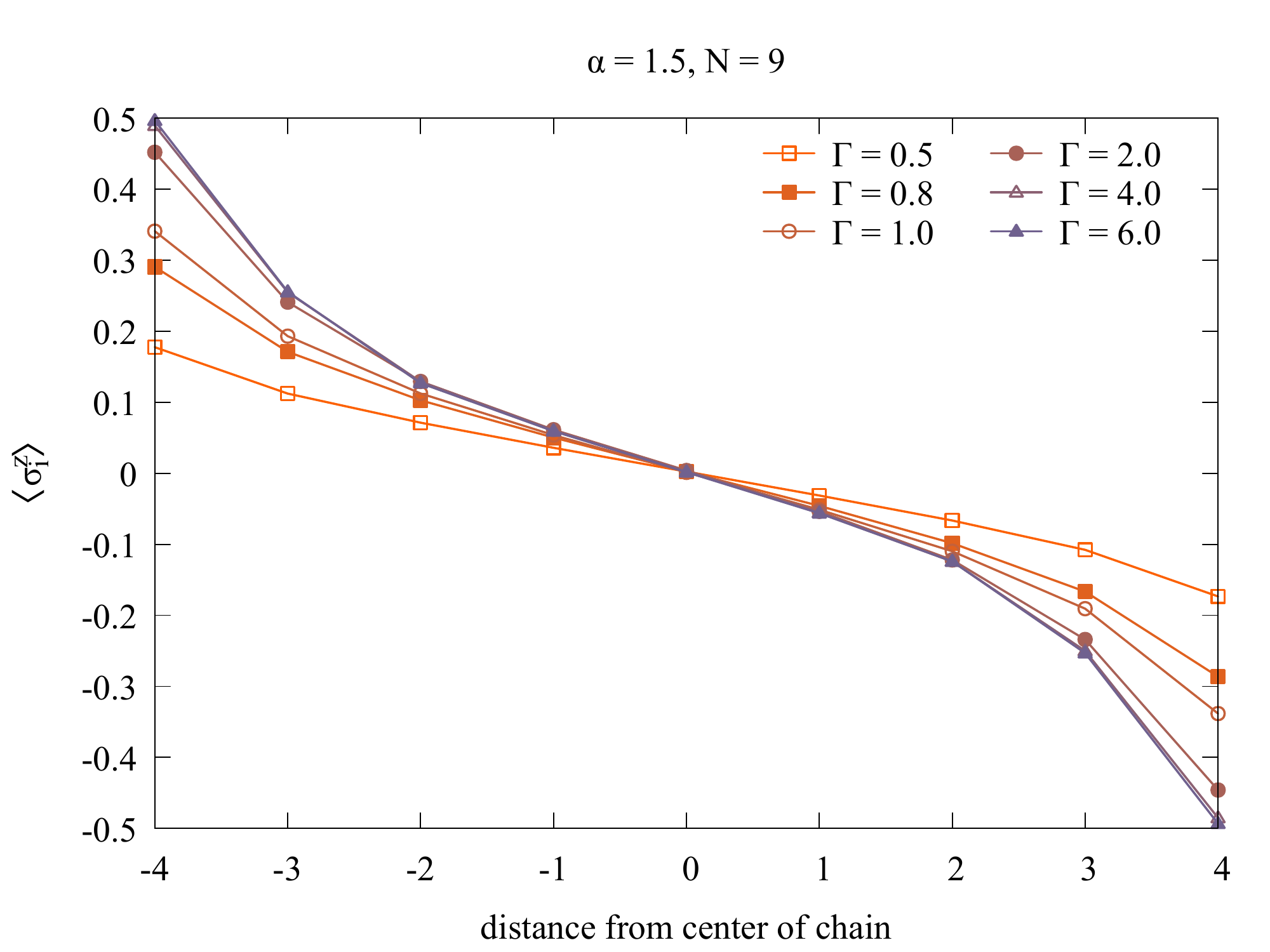}
	\caption{Polarisation profiles for different driving strengths $\Gamma$. The polarisation changes in value, but the distribution in the chain qualitatively stays the same over a large range of scattering rates. This explains why its impact on the\neu{transition length of the chain}is small.}
	\label{fig:A2}
\end{figure}

Fig.~\ref{fig:A1} shows polarisation plots for different chain lengths for two different long-range scenarios, with $\alpha=0.5$ and $\alpha=1.5$, respectively. The value of the\neu{bottleneck at its saturation}is smaller with less long-range, and it is reached for a much longer\neu{transition chain length.}\\
\\
Fig.~\ref{fig:A2} shows polarisation profiles for a chain of length $N=9$ for different driving scenarios $\Gamma$. We see that although the quantitative value of the transport bottleneck is changed by the driving, the qualitative picture of the polarisation distribution is not really changed. This explains why the transition to the ballistic regime is relatively independent of the driving strength.
\section{Details on the quantum trajectories method}\label{qmcappendix}
\begin{figure}[t!]
	\includegraphics[width=\linewidth]{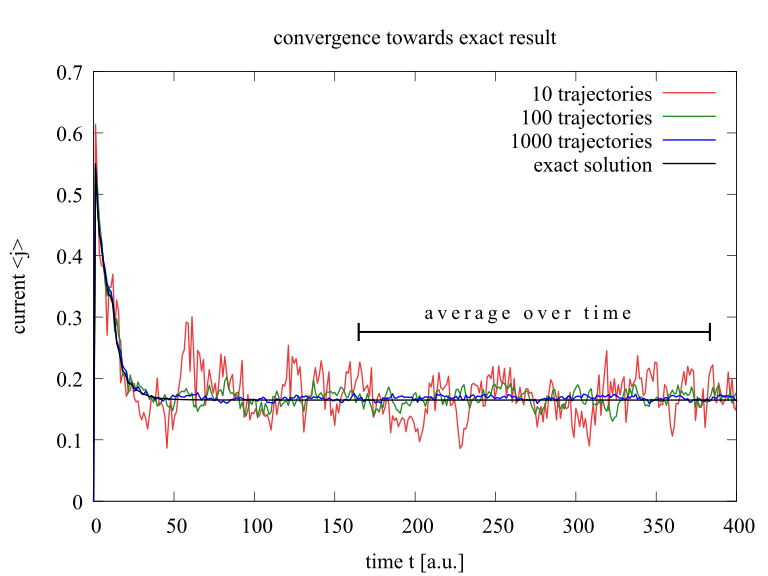}
	\caption{\neu{Convergence towards the exact solution of the master equation for a chain of N = 7 sites, with $\alpha=2.0$. The standard deviation is $s_{10}\approx0.0357$, $s_{100}\approx0.0114$, $s_{1000}\approx0.0036$, respectively. Note that we obtain all steady state values in this work by an additional averaging over many points in time, which further increases the precision by far.}}
	\label{fig:A3}
\end{figure}

\neu{The numerical approach we apply is relatively common in the field of open quantum systems, and was thoroughly described before~\cite{plenioknightrmp98,daleyreview2014trajectories}. The basic idea is to reduce the computational effort in the simulation of a dissipative system which cannot be described directly by just solving the Schrödinger equation, which is non-dissipative by design, due to the incoherent driving via Lindblad-Superoperators, or more precisely the so called jump-terms of the form $\sigma^{\pm}\rho\sigma^{\mp}$. This problem is  typically solved by a description in the master equation picture, which makes it necessary to calculate the squared Hilbert space of dimension $2^{2N}$. The idea behind the quantum trajectories method is to compute the dynamics with a non-hermitian effective Hamiltonian $H_\text{eff} = H_\text{sys} + \sqrt{\Gamma}\sigma^{\pm}\sigma^{\mp}$ on the level of the Schrödinger equation, and treat the mentioned jump terms separately. Before every timestep, the probability for a jump is evaluated $ \delta p = \Delta t \kappa\bra{\psi(t)}\sigma^{\pm}\sigma^{\mp} \ket{\psi(t)}$. Then, a random number $r\in[0,1]$ is pulled to decide whether a jump occurs or not. If $r<\delta p$, a jump occurs: $\ket{\psi(t+\Delta t)}=\frac{\sigma^\pm\ket{\psi(t)}}{\sqrt{\delta p/\kappa\Delta t}}.$ If $r>\delta p$ no jump takes place, therefore, the system evolves under the influence of the non-Hermitian effective Hamiltonian $\ket{\psi(t+\Delta t)}=\frac{(1-i\Delta tH_{\text{eff}}/\hbar)\ket{\psi(t)}}{\sqrt{1-\delta p}}.$ This process is repeated for every timestep to obtain a trajectory, and the average over many trajectories yields the desired observables.}This way the computation of the squared Hilbert space can be avoided, which reduces the number of equations to $2^N$.
In our case, the number of iterations necessary for a fairly accurate average seldom reaches 1000 realizations, and, more importantly, \textit{this number does not significantly change with growing N}. \neu{The master equation can be reproduced to very good accuracy, cf.~Fig.~\ref{fig:A3}. Repeating the convergence test each 100 times gave standard deviations of $s_{10}\approx0.0357$, $s_{100}\approx0.0114$, $s_{1000}\approx0.0036$, for the case of 10, 100 and 1000 trajectories, respectively. (The values have to be interpreted by comparison with the steady state current, which for this example is $\langle j \rangle = 0.1674$.)}The fact that within the Schrödinger equation, the Hilbert space scales only with $2^N$, leads to a clear numerical advantage over solving the master equation, at the latest from chain lengths of N=8 onwards. Thus, this method opens the possibility to examine chain lengths up to $N\approx20$, while for the same setting, the limit for the master equation approach seems to be somewhere not too far over $N=10$. Note that the t-DMRG methods, which are able to simulate chains of hundreds of spins, are less apt here, since this is an open chain, which means it has to be treated in the Liouville space, and additionally, the long-range interaction can not very effectively be simulated in an algorithm specifically designed for nearest-neighbour coupling~\cite{DaleyMPSReview}. It furthermore leads to matrix product states which are not sparse enough to make truncation outweigh the numerical cost of the algorithm, especially not in the far from equilibrium regime of very strong driving.%
Note that unlike e.g. for master equation simulations, in our framework it is \textit{impossible by design, to let the algorithm find a steady state} and then automatically end the simulation. The points in time where the jumps occur are completely unpredictable, and the single trajectory does not give any hints on whether a steady state is close or far. This makes the analysis of the transient behaviour of the averaged trajectories a crucial task when examining the convergence. There are e.g. cases where there is a late intersection taking place between the transient currents of chains of different lenghts N. It is important to certify that the simulation time was long enough to cover those intersections, since otherwise one runs into the danger of drawing qualitatively wrong conclusions.
Besides, the fact that the single QMC-trajectory does not know a steady state also leads to the fact that unlike in the master equation case, it is \textit{not advisable to initiate the spin chain in an initial state close to the steady state}. It proved a lot more efficient to start the dynamics in a $\ket{\psi(t=0)}=\ket{\uparrow\downarrow\downarrow...}$ initial state.%

Note that the \textit{transient dynamics will become significantly longer with growing chain length} (also in the master equation framework of course). This has to be taken into account when extrapolating the needed simulation time for long chains from shorter ones, since you will certainly not reach steady state values for the same simulation times. (Increasing N by one in our case leads to a transient around 200-300 time units longer).
Even for a comparatively big number of iterations, the probability to exactly hit the 'real' value of the steady state precisely at the end of the simulation is not very high. To address this, it proved very effective to allow the simulation be long enough that it shows a sufficiently long time interval at the end where it does no longer change its average value. This last part of the simulation time can then serve as a basis for an \textit{averaging over time},\neu{cf.~Fig.~\ref{fig:A3}. Due to the relatively long transients, especially for long chains, }the necessary extra time does not produce a new bottleneck in the computation time. On the contrary, this little trick leads to a drastic reduction of iterations necessary to reproduce the master equation dynamics and is thus strongly recommended.%

Another particularity of Monte Carlo is its sensitivity towards the \textit{simulation time step}. We observe that there is an optimal timestep, and that for even smaller dt, the simulations become more unstable again. This is counter intuitive, since from other numerical methods we are used to a rising stability with smaller step sizes. This issue was already reported in 1993~\cite{Molmer93}, apparently there is a general lower limit of the time-step within the Monte Carlo framework, an issue those authors argue to be rooted in the time-scale approximations made when deriving the method. They state that for too small dt, the correlation times of the bath seem to begin to play a role again, although they where previously omitted.%
\bibliography{Bib}
%

%
\end{document}